\documentclass[conference]{IEEEtran}
\IEEEoverridecommandlockouts

\usepackage{url}
\usepackage{cite}
\usepackage{amsmath,amssymb,amsfonts}
\usepackage{algorithmic}
\usepackage{graphicx}
\usepackage{textcomp}
\usepackage{xcolor}
\def\BibTeX{{\rm B\kern-.05em{\sc i\kern-.025em b}\kern-.08em
    T\kern-.1667em\lower.7ex\hbox{E}\kern-.125emX}}

\begin{document}

\title{A Real-time DDS-Based Chest X-Ray Decision Support System for Resource-constrained Clinics}

\author{\IEEEauthorblockN{Omar H. Khater}
\IEEEauthorblockA{\textit{Department of Computer Engineering} \\
\textit{King Fahd University of Petroleum \& Minerals}\\
Dhahran, Saudi Arabia \\
g202313250@kfupm.edu.sa}
\\
\IEEEauthorblockN{Farouq Aliyu}
\IEEEauthorblockA{\textit{ARC for Nonprofit and Social Development} \\
\textit{King Fahd University of Petroleum \& Minerals}\\
Dhahran, Saudi Arabia \\
farouq.muhammad@kfupm.edu.sa}
\and
\IEEEauthorblockN{Basem Almadani}
\IEEEauthorblockA{\textit{ARC for Nonprofit and Social Development} \\
\textit{King Fahd University of Petroleum \& Minerals}\\
Dhahran, Saudi Arabia \\
mbasem@kfupm.edu.sa}
\\
\IEEEauthorblockN{Esam Al-Nahari}
\IEEEauthorblockA{\textit{ARC for Nonprofit and Social Development} \\
\textit{King Fahd University of Petroleum \& Minerals}\\
Dhahran, Saudi Arabia \\
alnahari@kfupm.edu.sa}
}

\maketitle


\begin{abstract}
Internet of Things (IoT)-based healthcare systems offer significant potential to improve the delivery of healthcare services in humanitarian engineering, providing essential care to millions of underserved people in remote areas worldwide. However, these areas have poor network infrastructure, making communications difficult for traditional IoT systems. This paper presents a real-time chest X-ray classification system for hospitals in remote areas. The system uses a ResNet50 deep learning model for disease classification and FastDDS real-time middleware for reliable communication between the health practitioners and the model. We fine-tuned the ResNet50 neural network to achieve 88.61\% accuracy, 88.76\% precision, and 88.49\% recall. Our system results gain an average throughput of 3.2 KB/s and an average latency of 65 ms. The proposed system demonstrates how middleware-based systems can assist doctors in remote locations.
\end{abstract}


\maketitle

\section{Introduction}
The global population continues to grow rapidly, driving healthcare demands that increasingly outpace the capacity of healthcare systems worldwide and exacerbating shortages of medical professionals in many regions. A recent study based on World Health Organization (WHO) and United Nations (UN) data from 2018 to 2020 reports that although more than 13 million physicians are available globally, corresponding to an average of 19.5 doctors per 10,000 people, their distribution is highly uneven \cite{Kharazmi2024}. Approximately 43\% of physicians are concentrated in high Human Development Index (HDI) countries, which serve only 20\% of the global population, while low HDI regions have access to just 1\% of the worldwide healthcare workforce. 

This imbalance places significant strain on healthcare systems and underscores the urgent need for strategic interventions. The COVID-19 pandemic further exposed these vulnerabilities, severely overwhelming healthcare infrastructures and highlighting the critical shortage of specialized medical staff, particularly chest physicians. These challenges emphasize the importance of developing reliable, fully automated diagnostic support systems that can assist clinicians, reduce workload, and accelerate the diagnostic process, especially in resource-constrained settings.

Recent vision-based deep learning models have demonstrated their efficiency in analyzing medical images, extracting critical features from X-ray images, and predicting chest diseases such as COVID-19 \cite{pereira2020covid}. Convolutional Neural Networks (CNNs) have proven reliable for diagnosis \cite{santosh2020truncated}. ResNet50 proved its ability to classify the X-ray images. It can learn complex features and capture patterns due to its skipped connections, which enable more efficient training of deep network layers \cite{koonce2021resnet}. Additionally, it avoids vanishing gradients through the residual connections. However, although deep learning models' accuracy ensures the quality of insights, robust communication performance ensures these insights are delivered timely and reliably, which is paramount in dynamic and critical healthcare environments \cite{almadani2025systematic}.

\begin{figure}
    \centering
    \includegraphics[width=1.0\linewidth]{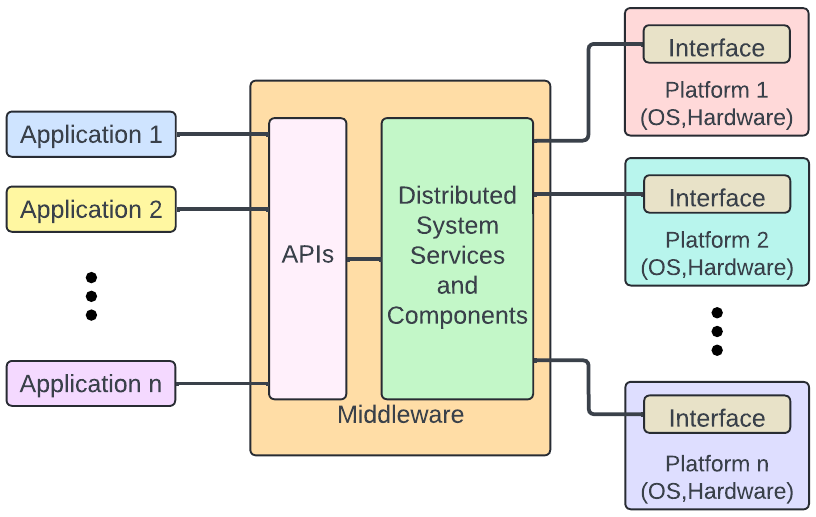}
    \caption{A Typical Middleware}
    \label{fig:middleware}
\end{figure}

Therefore, a middleware is a crucial component of modern distributed computing and data management.
As shown in Fig. \ref{fig:middleware}, middleware serves as an intermediary layer that connects operating systems and applications. It provides a standard interface that smoothly integrates the various system components. We chose FastDDS in this paper because of its scalability and high performance \cite{bode2023dds}. Additionally, FastDDS provides reliable, low-latency data transfer, which is essential for real-time communication. Moreover, it is open-source, compliant with the DDS standard, and applicable across different platforms \cite{zhang2021middleware}.

This paper developed an end-to-end deep learning–based decision-support system for distributed healthcare environments that assists clinicians in diagnosing chest diseases using X-ray images. The proposed system enables doctors to publish chest X-ray (CXR) data and receive automated diagnostic results through a scalable publish–subscribe architecture supported by real-time middleware. A Residual Network 50 (ResNet50) is employed to improve feature representation and classification accuracy, while middleware integration ensures reliable, low-latency communication between distributed system components, especially in a resource-constrained environment. The rest of the paper is organized as follows. Section \ref{sec:review} reviews middleware technologies and state-of-the-art vision-based deep learning models for chest disease diagnosis. Section \ref{sec:method} presents the proposed system, Section \ref{sec:exp} describes the experimental setup, Section \ref{sec:result} discusses the results, and Section \ref{sec:conc} concludes the paper and outlines future work.


\section{Literature Review}
\label{sec:review}

Several research show that deep learning can diagnose chest-related diseases using X-ray. For example, Alshmrani et al. \cite{alshmrani2023deep} presented a deep learning architecture for multi-class lung disease classification using CXR images. The authors applied transfer learning on a pre-trained VGG19 network. The model was evaluated on a large-scale dataset comprising over 21,000 labeled CXR images spanning multiple classes, including COVID-19, pneumonia, lung opacity, viral pneumonia, and normal cases. Experimental results demonstrated strong classification performance, achieving an accuracy of 96.48\%, highlighting the effectiveness of hybrid deep learning architectures for automated chest disease diagnosis.

In a related study, Karaddi and Sharma \cite{karaddi2023automated} evaluated several pre-trained convolutional neural networks, including AlexNet, DenseNet-201, ResNet-18, and InceptionResNetV2, for the classification of COVID-19, pneumonia, pneumothorax, tuberculosis, and normal cases using CXR images. Their experiments found that DenseNet-201 achieved the best performance, reaching an accuracy of up to 97.49\%. Recognizing the importance of an end-to-end decision-support system for CXR-based diagnosis, the authors proposed developing a deployable Android and Windows application that leverages a cloud-based lung disease classification model in their future work. However, such cloud-centric architectures are not well-suited for safety-critical healthcare applications in remote and underserved areas, where network connectivity is often unreliable, and bandwidth costs are prohibitive. Consequently, deploying diagnostic systems over local or wide-area networks (LAN/WAN) with real-time communication support is more appropriate for ensuring timely and reliable clinical decision-making.

Recent studies show that middleware technologies can handle the growing complexity of real-time communication and data processing across a variety of fields, including distributed systems, robotics, and healthcare \cite{almadani2025systematic,robotics2025gambo}. Depending on their design and intended purpose, these technologies provide different trade-offs in terms of performance, scalability, simplicity, and real-time assurances: Fore example, Snout \cite{becker2022snout} is a lightweight middleware designed to simplify interaction with software-defined radios (SDRs). It provides a minimal framework that requires limited technical expertise and is optimized for low memory usage and reduced CPU overhead, making it suitable for real-time applications in resource-constrained environments such as embedded systems and small networked devices. However, this simplicity comes at the cost of limited scalability, as Snout is not well suited for large-scale or geographically distributed systems where communication demands can exceed its capacity and require more sophisticated data management. In contrast, middleware solutions based on the Data Distribution Service (DDS) standard are specifically designed to address the challenges of real-time data exchange in distributed systems, offering greater scalability and robustness for complex communication scenarios.

FastDDS, CycloneDDS, RTI Connext, and OpenDDS are examples of middleware implementations based on the Data Distribution Service (DDS) standard, designed for distributed real-time communication among heterogeneous devices. These systems are well-suited for large-scale and safety-critical applications such as industrial control systems, autonomous vehicles, and healthcare systems, as they efficiently manage data exchange across multiple nodes. Bode et al. \cite{bode2023dds} report that FastDDS and CycloneDDS provide a strong foundation for real-time distributed systems by achieving lower latency and reduced packet delay variation compared to earlier DDS implementations. Unlike lightweight middleware such as Snout, which targets localized and low-overhead use cases, DDS middleware scales effectively across wide-area and complex network deployments. Nevertheless, because DDS implementations typically rely on the Operating System's (OS) network stack, which is not inherently optimized for real-time performance, achieving strict real-time guarantees remains challenging. This trade-off highlights that while DDS middleware introduces greater complexity, it is better suited to demanding distributed applications, like a distributed healthcare system, whereas Snout prioritizes simplicity and efficiency in small-scale scenarios.

\begin{table*}[!ht]
\centering
\caption{Comparison of Real-Time Middleware Technologies for Distributed Healthcare Applications}
\label{tab:middleware_comparison}
\renewcommand{\arraystretch}{1.2}
\setlength{\tabcolsep}{4pt}
\begin{tabular}{ c p{1.75cm} p{2.0cm} c c p{3.2cm} p{3.2cm} c }
\hline
\textbf{Ref.} & \textbf{Middleware} & \textbf{Target Domain} & \textbf{Scalability} & \textbf{Complexity} & \textbf{Strengths} & \textbf{Limitations} & \textbf{Healthcare} \\
 &  &  &  &  &  &  & \textbf{Suitability} \\
\hline
\cite{becker2022snout} & Snout & Embedded / SDR systems & Low & Low & Lightweight, low CPU and memory overhead, simple deployment \newline & Limited scalability, weak support for distributed systems \newline & No \\
\cite{bode2023dds} & FastDDS & Distributed real-time systems & High & Moderate & Low latency, scalable publish--subscribe, cross-platform, open-source \newline & Dependence on OS network stack \newline & Yes \\
\cite{bode2023dds} & CycloneDDS & Distributed real-time systems & High & Moderate & Reduced latency and packet delay variation, DDS compliant \newline & OS network stack dependency \newline & Yes \\
\cite{bode2024adopting} & DDS + \newline DPDK/XDP & High-performance real-time systems \newline & High & High & Significant latency reduction, high throughput \newline & High setup and maintenance complexity \newline & Partial \\
\cite{dalkiran2021automated} & JMS--DDS  \newline Hybrid & Hybrid real-time / non-real-time systems \newline & High & High & Separation of time-critical and non-critical communication \newline & Java dependency, reduced portability \newline & Partial \\
\cite{laurenzi2023xbot2} & XBot2 & Robotics  \newline real-time systems \newline & Moderate & High & Deterministic execution, strong real-time guarantees \newline & Linux-only, limited interoperability \newline & No \\
\cite{peeck2021middleware} & DDS with \newline Retransmission Enhancements & Wireless and unstable networks & High & High & Prioritized retransmission, resilience to packet loss \newline & Added system complexity \newline & Yes \\
\hline
\end{tabular}
\end{table*}


Researchers have investigated integrating user-space networking technologies, such as Express Data Path (XDP) and Data Plane Development Kit (DPDK), to improve the performance of DDS middleware and overcome real-time performance limitations. Bode et al. \cite{bode2024adopting} showed that using CycloneDDS in conjunction with DPDK significantly reduces latency and increases throughput, resolving some of the real-time issues with conventional DDS implementations. They found that a CycloneDDS-DPDK-based system reduces mean latency by up to 31\%, making it an attractive option for high-performance, networked real-time systems like large-scale robotic systems or high-frequency trading. However, DPDK's technical complexity, including setup and maintenance difficulties, could lead to high downtime, which is not ideal in healthcare systems. In contrast, Snout provides a lower barrier to entry due to its simpler setup, despite scalability constraints, especially for smaller applications that do not require the intensive performance improvements offered by DPDK.

Systems that integrate real-time and non-real-time components, such as military applications where Combat Management Systems (CMS) must interface with Command and Control (C2) systems, or healthcare environments where diagnostic decision-support systems interact with hospital information and billing platforms, introduce additional middleware complexity. Dalkıran et al. \cite{dalkiran2021automated} investigated a JMS–DDS hybrid architecture that combines the Data Distribution Service (DDS) with the Java Message Service (JMS) to enable communication between time-critical and non–time-critical system components. In this approach, JMS provides flexible asynchronous messaging for non-real-time tasks, while DDS supports deterministic communication for real-time data exchange. This separation allows complex systems to be integrated while preserving real-time behavior for critical components, which is essential in safety-critical domains such as military and healthcare environments where coordination and responsiveness are paramount. However, reliance on the Java ecosystem introduces limitations in terms of portability and deployment flexibility. As a result, JMS–DDS architectures are particularly suited for hybrid real-time systems, whereas DDS-only solutions such as FastDDS and CycloneDDS remain more platform-neutral but offer less native support for integrating non-real-time messaging components.

Middleware also plays a critical role in safety-critical real-time domains such as robotics, telesurgery, and Intensive Care Unit (ICU) monitoring and control, where systems must coordinate multiple hardware and software components under strict timing constraints. Laurenzi et al. \cite{laurenzi2023xbot2} introduced XBot2, a middleware framework designed for multi-threaded real-time robotic systems. Compared to general-purpose platforms such as the Robot Operating System (ROS), which offer limited real-time guarantees and face challenges with intra-process communication, XBot2 provides improved determinism and modularity through its hardware abstraction layer and real-time execution model. Although such domain-specific middleware demonstrates strong performance in tightly coupled, real-time environments, its reliance on Linux-based platforms limits its portability across heterogeneous, distributed systems using other OSs. In contrast, DDS-based middleware such as FastDDS prioritizes cross-platform compatibility and scalable distributed communication, making it more suitable for applications like distributed medical decision-support systems, where interoperability, scalability, and deployment flexibility are essential.

\begin{figure*}[!ht]
 \centering
 \includegraphics[scale=0.5]{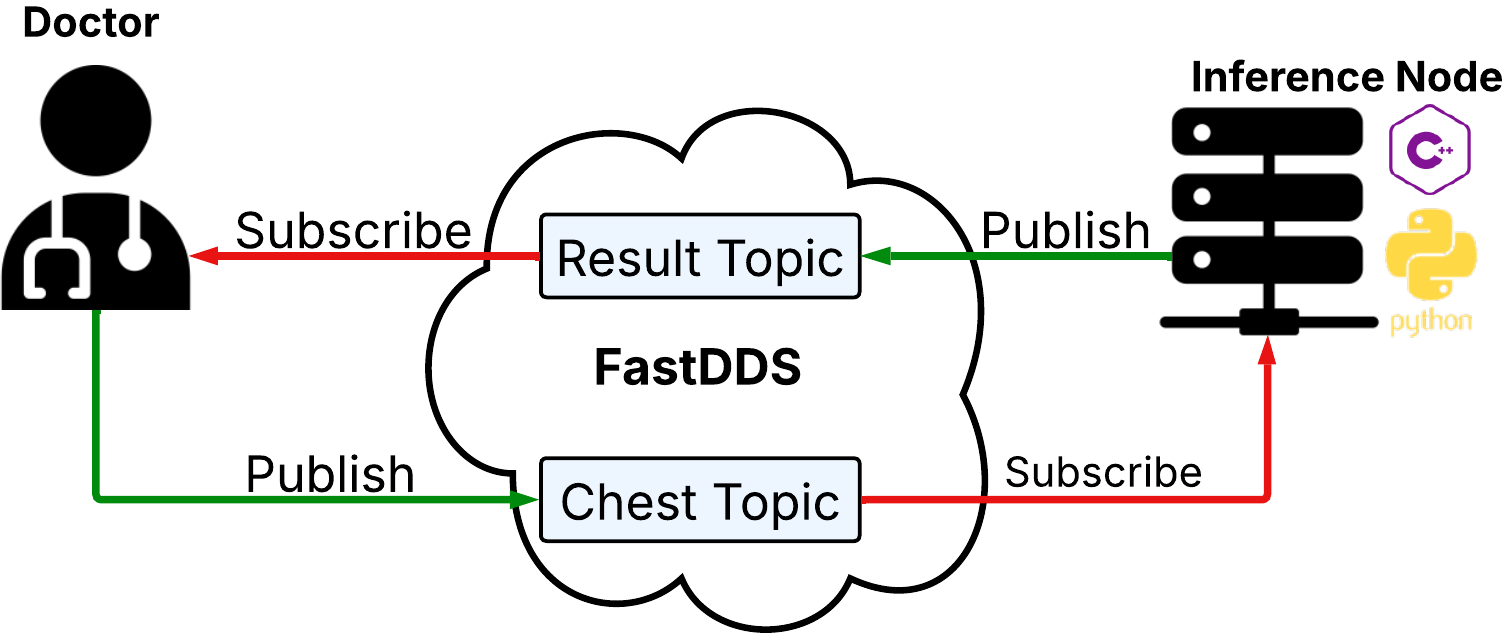}
 \caption{The proposed system.}
 \label{fig:example}
 \end{figure*}

Additionally, middleware plays a critical role in wireless communication systems that require real-time data transfer over unstable networks, such as telemedicine applications in remote areas where network instability and packet loss are common. To address these challenges, Peeck et al. \cite{peeck2021middleware} enhanced DDS middleware by introducing a retransmission strategy that prioritizes data based on its temporal significance. This approach provides an advantage over conventional transport protocols such as TCP and UDP by improving the delivery of time-sensitive data under lossy network conditions. While FastDDS and other DDS implementations offer reliable communication for distributed systems, their performance can still degrade in highly unstable networks, motivating the need for such enhancements. This further highlights the trade-off between simplicity and the ability to manage complex network environments, as lightweight solutions such as Snout lack the mechanisms required to support large-scale distributed communication and prioritized data retransmission.

As summarized in Table~\ref{tab:middleware_comparison}, FastDDS provides the most balanced trade-off among real-time support, scalability, deployment complexity, and interoperability for distributed healthcare applications. Unlike lightweight middleware such as Snout, which offers real-time capabilities but lacks scalability, FastDDS is designed to support large-scale distributed systems while maintaining low-latency communication. Compared to high-performance DDS configurations enhanced with DPDK or XDP, FastDDS avoids excessive setup and maintenance complexity, which is critical for deployment in resource-constrained and underserved healthcare environments. Additionally, while hybrid solutions such as JMS–DDS enable integration of real-time and non-real-time components, their dependence on the Java ecosystem reduces portability and increases system complexity. Domain-specific middleware such as XBot2, although capable of strong real-time performance, is limited to tightly coupled robotic systems and lacks cross-platform interoperability. Consequently, FastDDS emerges as the most suitable middleware for the proposed distributed medical decision-support system, offering scalable, real-time communication with manageable complexity and broad platform support.


\section{Proposed System}
\label{sec:method}

This section presents the proposed real-time chest X-ray decision-support system and describes how its components interact to provide reliable, low-latency diagnostic support in resource-constrained healthcare environments. The overall system architecture is first introduced, followed by the integration of the FastDDS middleware for real-time communication. Finally, the inference node and the deep learning model used for automated disease classification are explained.

Fig. \ref{fig:example} illustrates the proposed distributed decision-support system, which integrates the FastDDS publish–subscribe middleware with a vision-based deep learning model to assist clinicians and accelerate the CXR diagnosis process. The system follows a distributed architecture in which multiple clinicians can access one or more deep learning inference nodes for decision support, while middleware enables reliable and low-latency communication between system components.

The proposed system consists of two main nodes connected through the FastDDS middleware. The first node is the doctor node, which is responsible for publishing CXR images to the middleware and subscribing to the corresponding diagnostic results. To emulate a resource-constrained Internet of Things (IoT) device, such as a smartphone, this node is implemented on a Raspberry Pi. The second node is the inference node, which subscribes to the published X-ray images, performs automated inference using a trained deep learning model, and publishes the diagnostic results back to the middleware. A laptop is used for the inference node to represent a device with higher computational capability than the IoT node, while remaining less powerful than a cloud-based server.

The FastDDS middleware manages data exchange between publisher and subscriber nodes, ensuring efficient, scalable communication across the distributed system. All FastDDS publisher and subscriber components in this work are implemented using the C++ programming language. This design choice is motivated by the need for low-latency and predictable communication, as the native FastDDS API is optimized for C++ and provides finer control over memory management, threading, and Quality of Service (QoS) policies compared to higher-level language bindings. Using C++ also reduces runtime overhead, which is important for real-time data exchange in distributed healthcare environments. We also used the default QoS policies settings.

\begin{figure}[!ht]
    \centering
    \includegraphics[scale=0.5]{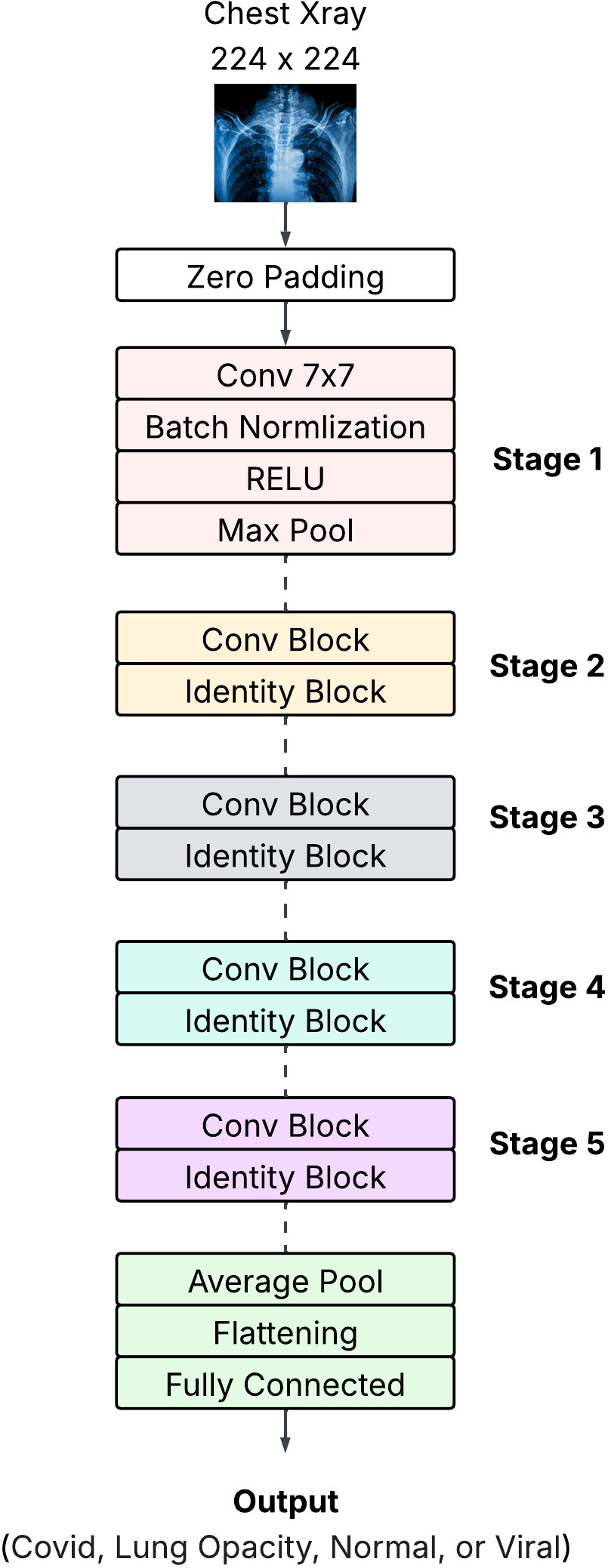}
    \caption{Deep Learning Model for Inference Node}
    \label{fig:model}
\end{figure}

\begin{table}[!ht]
\caption{Layers of the Proposed ResNet50}
\label{tab:resnet50}
\begin{tabular}{lllc} \hline
Section        & \# of Blocks & Layers per Block & Total Layers \\ \hline
Stem (Stage 1) & N/A              & 1 (Conv7x7)      & 1            \\
Stage 2        & 3 Blocks         & 3 Layers each    & 9            \\
Stage 3        & 4 Blocks         & 3 Layers each    & 12           \\
Stage 4        & 6 Blocks         & 3 Layers each    & 18           \\
Stage 5        & 3 Blocks         & 3 Layers each    & 9            \\
Output (FC)    & N/A              & 1 (Linear)       & 1            \\
GRAND TOTAL    &                  &                  & 50           \\ \hline
\end{tabular}
\end{table}

Upon receiving a CXR image from the doctor node, the inference node automatically triggers the inference pipeline. Specifically, the subscriber application implemented in C++ invokes a Python-based inference script that executes the trained ResNet50 model to perform diagnosis. This approach enables seamless integration between the real-time FastDDS communication layer and the deep learning inference logic. We used the default ResNet50 implementation form PyTorch \cite{PyTorch_2017}. Fig. \ref{fig:model} shows the ResNet50 model in our proposed system. The input layer take a 224$\times$224 CXR image in JPEG format. Table \ref{tab:resnet50} shows the layers in each stage of the model. The dotted lines are the ResNet Skip Connections that allow the input signal to skip over the weighted layers (the convolutions) and be added directly to the output of that block. We modify the fully connected output layer to predict the four target classes (i.e., Covid, Lung Opacity, Normal, or Viral). Once inference is complete, the diagnostic result is published back to the FastDDS middleware and delivered to the doctor node, which subscribes to and retrieves it, thereby completing the end-to-end decision-support workflow.


\section{Experimental Setup}
\label{sec:exp}
The experiment setup for the proposed system consists of a Raspberry Pi and a laptop PC communicating over Wi-Fi using FastDDS. The Raspberry Pi runs the Linux operating system, and it has 4 GB of RAM. The laptop PC comprises an NVIDIA GeForce RTX 3080 Ti with 16 GB of memory, and the RAM is 32 GB.

The dataset used in this work consists of 21,269 labeled CXR images spanning four classes and was employed to fine-tune the ResNet50 model for chest disease classification \cite{rahman2021exploring, chowdhury2020can}. The dataset was obtained from Kaggle, an open-source repository, and includes X-ray images of COVID-19, normal chest, lung opacity, and viral pneumonia cases. Specifically, the dataset comprises 3,617 COVID-19 images, 10,193 normal chest images, 6,013 lung opacity images, and 1,346 viral pneumonia images. For model development and evaluation, the dataset was partitioned into 80\% for training, 10\% for validation, and 10\% for testing.

The model was trained for 30 epochs with full fine-tuning: unlike conventional transfer learning approaches that freeze the backbone layers, all network parameters were allowed to be updated. The ResNet50 architecture was initialized with ImageNet-pretrained weights provided by the PyTorch library \cite{PyTorch_2017a}, and the entire network was fine-tuned on the CXR dataset to adapt the learned feature representations to the specific characteristics of medical radiographs.

\section{Results and Discussions}
\label{sec:result}

\begin{figure}[!ht]
    \centering
    \includegraphics[width=1.0\columnwidth]{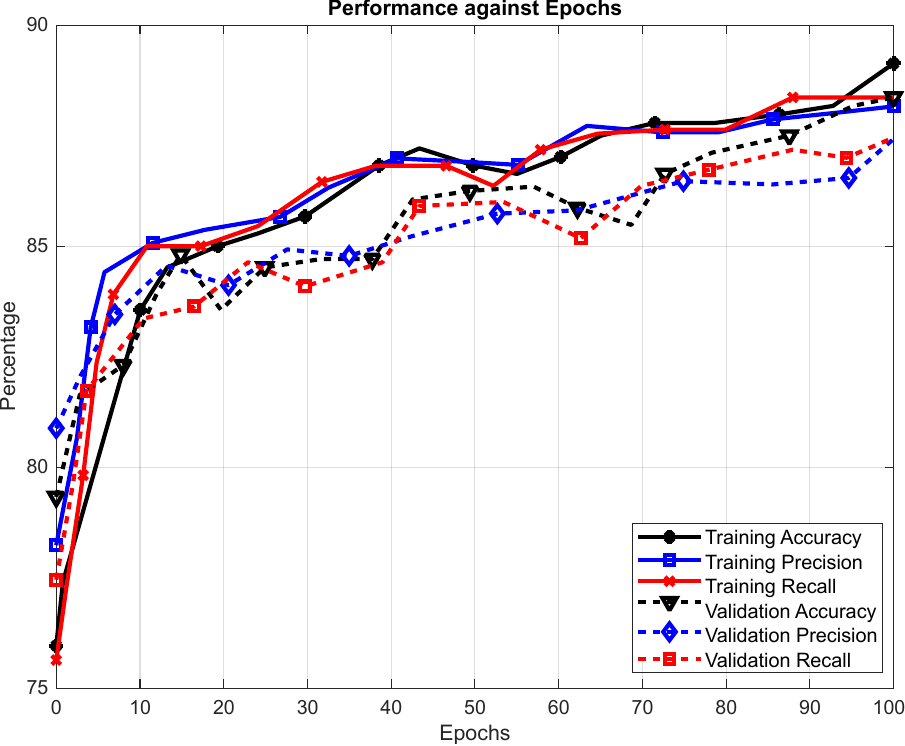}
    \caption{Performance of ResNet50}
    \label{fig:ResNet50_Performance}
\end{figure}

Fig.~\ref{fig:ResNet50_Performance} shows the performance of the proposed ResNet50 Model after fine-tuning it on the CXR image dataset. The proposed model achieved an accuracy of 88.61\%, precision of 88.76\%, and recall (sensitivity) of 88.49\%. These results are within the range of diagnostic performance reported for clinician-interpreted chest radiographs, as described in a recent systematic review and meta-analysis by Qafesha et al. \cite{Qafesha2025chest}, which evaluated human interpretation of CXR using high-resolution computed tomography as the reference standard and reported an overall diagnostic performance (AUC) of approximately 0.88, despite a lower pooled sensitivity of approximately 62\%. This comparison suggests that the proposed model may serve as a supportive tool for clinical decision-making, particularly in CXR-based screening workflows, rather than as a replacement for definitive diagnostic imaging.

\begin{figure}[!ht]
\centering
\includegraphics[width=1.0\columnwidth]{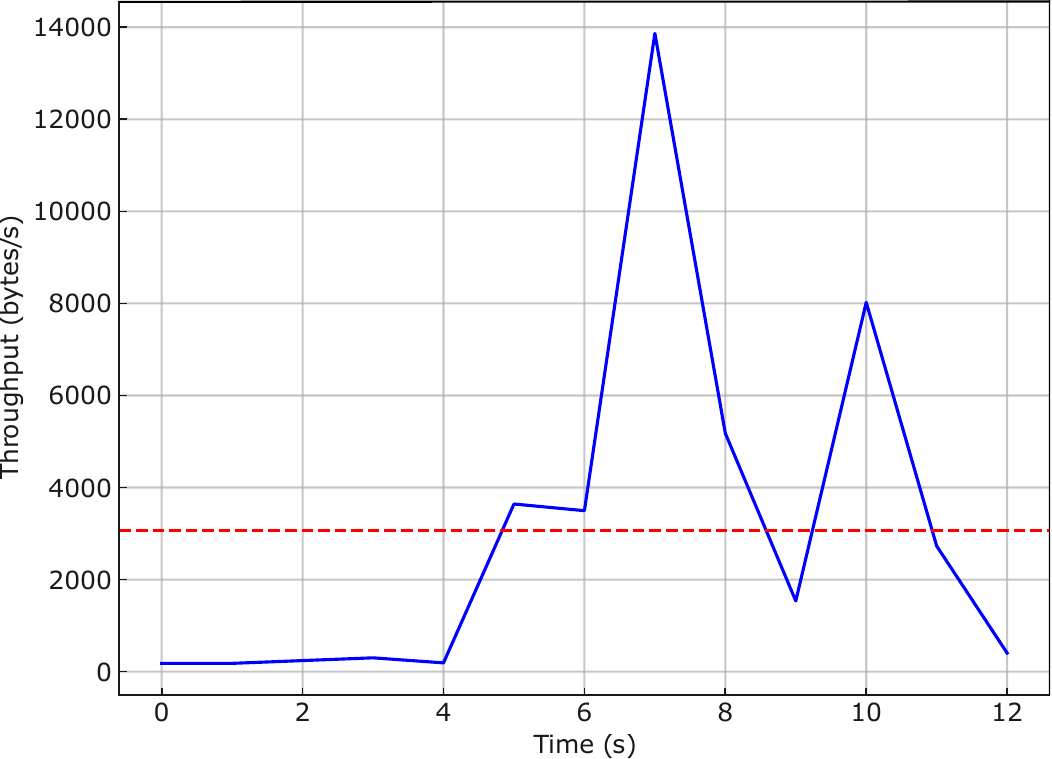} 
\caption{Throughput profile}
\label{fig: Throughput_profile}
\end{figure}

\begin{figure}[!ht]
\centering
\includegraphics[width=1.0\columnwidth]{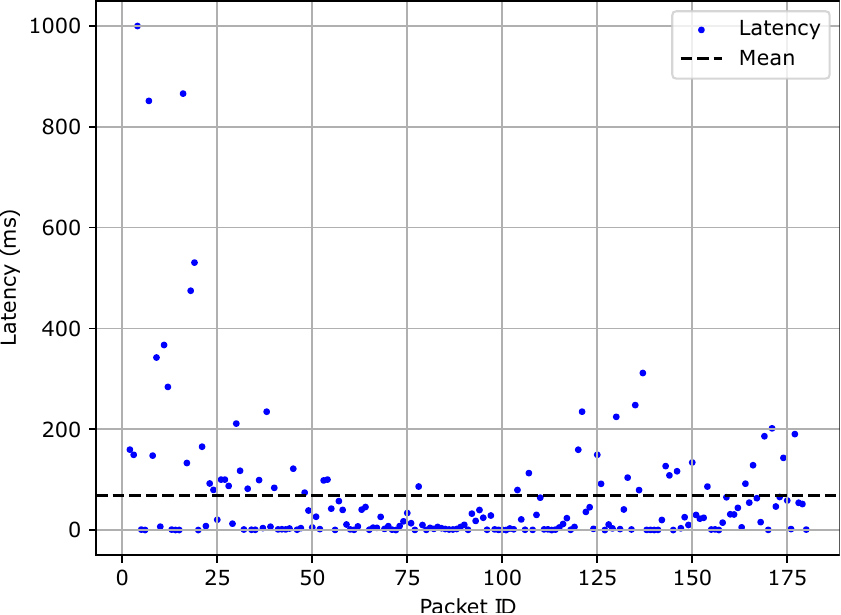} 
\caption{Latency profile}
\label{fig: Latency_profile}
\end{figure}

To investigate the communication performance of the system, we measured network latency and throughput over Wi-Fi. We analyzed the transmission latency from publishing the X-ray image on the doctor's node through FastDDS to the inference node, including the inference time, and ending with the inference node publishing the result through FastDDS and returning to the doctor's node. The results show an average latency of 65 ms. The whole transmission data was around 185 packets. Fig.~\ref{fig: Latency_profile} shows the visual representation of the transmitted packets' latency. The latency pattern provides a valuable insight into the operational characteristics of the communication between the doctor node and the inference node. Fig.~\ref{fig: Throughput_profile} shows that the throughput of the proposed system is around 3,200 bytes/second.

\section{Conclusion and Future Work}
\label{sec:conc}
This work investigated the integration of vision-based deep learning models with middleware to develop a system capable of assisting medical staff with chest disease classification. The ResNet50 model showed superior performance on X-ray images due to its ability to retain residual connections to prevent vanishing gradients and its efficient feature extraction, making it a suitable choice for the classification task. The model achieves 88.61\% accuracy, 88.76\% precision, and 88.49\% recall. Additionally, the FastDDS middleware is presenting outstanding performance in terms of throughput and latency. It enables the system to achieve an average throughput of 3,200 bytes/second and an average end-to-end latency of 65 ms.

In the future, the research should focus on improving scalability to ensure that doctors from different parts of the hospital or nearby clinics can use the system efficiently. Additionally, we will investigate system security using the FastDDS security module, as the system must be protected against attacks. Finally, we shall extend and add more inference nodes to provide fault tolerance, enable load balancing, and ensure the system performs efficiently across different scenarios.

\section*{Acknowledgment}
The authors would like to acknowledge all support provided by Alfozan Academy, the Applied Research Center for Nonprofit \& Social Development, and King Fahd University of Petroleum \& Minerals (KFUPM).

\bibliographystyle{ieeetr} 
\bibliography{reference} 

\end{document}